# Simulator as a Tool for the Future Maritime Education and Research: A Discussion

Yushan Pan, Arnfinn Oksavik and Hans Petter Hildre

Norwegian University of Science and Technology, Ålesund N-6025, Norway
`{yushan.pan;hans.p.hildre}ntnu.no,arnfinn.oksavik@ocean-training.no`

**Abstract.** A few studies in the maritime domain utilize co-design in ship design workshops, however, none of them addresses a full picture of how co-design can make changes in simulation-based maritime education. In this paper, we reflect how co-design can help to foresight future skills in the maritime domain, especially on how to use simulators to support increasing competence of seafarers and in turn to redesign simulators to support maritime education. Thus, we address collaborative and innovative research activities, to enable all participants (seafarers, trainers, technicians, authorities etc.) to share their experiences so a joint recognition of needed future skills can be reached. Along with the exchange of experiences, we assert that the supported simulations and simulator techniques could be designed to achieve sustainable growth for all participants.

**Keywords:** Simulation, Maritime Education, Maritime Research, Competence.

## 1   Introduction

As a result of improvements in information technology and infrastructure, maritime technologies and operations have dramatically evolved from traditional automatic, mechanical, mechatronic-based technologies to intelligence, human-centred, and information and communication technologies (ICT) supported smart operations. Such changes subvert the traditional evaluation of the competence of individual labours. At the same time, the difficulty of evaluating organisational usability[1] of those technologies is raised. The two are combined dramatically to challenge current maritime education and research in many levels.

In order to understand the problem of current maritime education and research, it is important to know how maritime competence is defined. The Ministry of Education and Research of Norway, which prioritizes non-cognitive skills and experience-based expertise [2], provides a definition: Competence is consistence of skills, knowledge, understanding, and attitude [3]. This means that if an individual wants to gain high competence in his or her field, simply knowing a lot of facts and rules, such as training procedures, provides only a basic understanding of the necessary skills. The person must also know how to find his own way around the knowledge needed in his profession [4].

Hence, it is noticeable that current simulator-based maritime education in Norway may not be able to offer a platform for seafarers to gain the highest level of competence if there is no suitable methodology. The reason for this is simple; land-based simulators are connected through a machine network to engage seafarers in the training process. Because technicians restrict this network to a predefined class of appropriate responses (cognitive skills of marine operators), the network incorporates the intelligence that was built into the machines by the technicians for that particular context. These skills reflect the competence of the technicians, not the competence of the seafarers. In addition, seafarers must follow the work procedures in their training programme. However, that is not true regarding seafarers' in-situ work practices at sea.

Why we argue competence? A brief history of maritime education in Norway helps. Norwegian maritime education consists of three main venues: upper secondary school and vocational education, technical colleges, and universities. Along with several training companies across the country, these three educational systems contribute to disequilibrium. For example, upper secondary school and vocational education, and technical colleges primarily focus on utilising simulators to train seafarers from the novice to the proficient level. After that, course certificates are awarded to students who later achieve some experience at sea then get certificates from the maritime authority of Norway. Certificates are primarily only paper that describes a position in the maritime industries. Alternatively training companies also offer training programs to seafarers and offer diplomas or certificates if the companies are approved by the Norwegian Maritime Authority [3]. On the other hand, comprehensive universities instruct technicians in how to design maritime simulations. There is no overlap between seafarers and technicians. In addition, technicians have less experience working at sea, while the seafarers have less knowledge of the simulators' capabilities and limitations. Altogether, the relationship between competences of individual labours and the above-mentioned missing links among organisations create a gap in which unavoidable fundamental questions are raised over the long term: Who has competence, who defines it, who evaluates it, and which relevant simulators are equivalent to in-situ knowledge and skills of which people in the work setting? This leads to an interesting research question: *How co-design as a research method contributes to the design of marine technologies, creating scenarios via simulators for example, in turn, to help designing simulation-based maritime education?*

## 2    Framing problems

As we knew that, competence cannot be transferred from one individual labour group to another simply through a fixed simulator. No one can duplicate the working experiences of others to produce the same success stories. However, only one thing can be learned from others: apply the lessons you learn to your daily work practice and obtain experiences to achieve competence. Thus, if a simulator is not able to function competently, why do we expect the formal training procedures that similar with 'formal mathematical or analytical rationality' [5] of the simulator to help seafarers gain a deep understanding of the competencies that build upon vast successful and non-duplicable experiences? Moreover, why do we only use the results from experimental work to misrepresent experience, another form of competence? Is it fruitful to help designing marine technologies with better and better scenarios? If knowledge bridges among different participants in the simulation-based maritime education is not built yet, then can we foresight future skills?

Gaining a high level of future skills in an unstructured area like maritime operations seems to require considerable concrete experiences with some type of structure. An individual person will be both an expert in certain types of methods in his or her own area of skill and less skilled in other areas. Being an expert, or being at any stage of skill acquisition, does not necessarily mean performing as well as everyone else who exhibits the same type of thought processes. Everyone function in at least one of five stages of skill level: novice, advanced beginner, competent, proficient, and expert [4]. A good proficient performer, such as a technician setting up a fixed simulator, while intuitively organising and understanding his task, will still find himself thinking analytically about what to do. The same applies to investigators of future maritime education. We have to admit that human competence is contrary to logic and reasoning.

How do we apply this understanding of the human learning process to the technology environment? How can we bring contributions from all participants to redesign technology (i.e., creating scenarios) and foresight future education? We must have a holistic understanding of the competence of seafarers, trainers, technicians, authorities, and managers and their simulator-supported interactive relationships toward decision-making. It is important to bear in mind, as scientists, that your users are not stupid [6] and that only the designed mechanism of training is, in most cases, the fault of scientists. Thus, co-design respects all users of simulators and can facility a design process for the maritime education. Probably it is not the only approach, but in our view, it is the best way to answer the question of who will evaluate whose competence through which joint agreement of what simulator competence.

## 3     Human learning and Competence

Looking at the maritime domain, upper secondary school and vocational education and technical colleges do train seafarers in gaining cognitive skills. However, cognitive skills are not full competence [3] and are rule-guided, expressed as "knowing that." If working situation is changed and thus requires new skills, a seafarer might not be able to handle it due to a lack of experience, expressed as "knowing how." This "knowing how" requires us to be broader participants to both build knowledge and exchange experiences. Together, we can build up an ecosystem to help develop competence and value for foresight future skills, including redesigning simulators to better support regulations and organizational restructuring.

It is noticeable that the distribution of maritime education is not the only thing that contributes to the gap. The International Convention on Standards of Training, Certification and Watch-keeping (STCW) for Seafarers [7] is also accountable. Notably, we do not admit that STCW has done something wrong. Instead, we illustrate that STCW has nothing to do with increasing seafarers' competence but only promises a procedure to train a novice seafarer and bring him or her to the proficient level. In addition, all these levels obey three principles [5] that help describe how things work: the practical level, the component level, and the functional level. These three principles follow basic rules and laws of physics and mathematics. For example, the simulator divides a particular job at sea into different components, each with its own function, and puts them all together to produce a result. This way, mechanistic functions are combined to encompass the functioning of the whole. Such top-down, context-independent analytical methods for cognitive skills are adopted to analyse competences of seafarers along a wide range between novice and proficient. For example, using a survey, questionnaire, and tools, we can evaluate human performance in simulators repeatedly until we get a satisfactory result.

The point is that no one can prove how many evaluations are enough because controlled experiments are not able to predict which unpredictable phenomena will cause failures. If we cannot manage what we choose to measure, we will not be able to control the cost of running experiments and will only create digital waste in most cases. All this will disable us from forecasting the usefulness of future skills for seafarers, trainers, technicians, authorities, and managers. As we are able to foresee and devise regulations for selecting future seafarers, it is important to address the transferring of competence through updated simulators. On one hand, we have to deal with participation, competence reuse, and competence transfer, while on the other hand, and decide how to combine these elements to shape simulator development.

## 4  Simulator-based Maritime Education and Research

Our experience is based on project examples taking place in the Norwegian maritime domain. The study should be viewed as a contribution to the ongoing debate on new methods, addressing how a planned approach can be successively and pragmatically modified and applied to foresight future skills in the maritime education, both for training seafarers and educating technicians, mangers and trainers. In many respects, the related work in similar maritime settings points to similar challenges. It has been difficult practically to involve the participants over time using the traditional engineering design approach, the same difficulty applies to the co-design approach too. Therefore, in the cross-sector setting involving semi-professionals in collaboration with the universities, we have to choose carefully who will be the participants and the possibility to be engaged to integrate the new collaborations and include the development of simulations through the experience at sea.

However, there is a long way to go since a well-established research group is needed. Historical issues caused the gaps of maritime education in higher education and have already leading to unsystematic structure for research and development of maritime technology. Well, the most bogeyman problem is to unfruitfully picture an incomplete work practices of participants. For example, Mallam et al. (2017) designed an 'ergonomic ship-evaluation tool' for introducing participatory design as a method to design a ship. The tool can create an environment that will help naval architects, crews and ergonomists work together to develop human-centred design solutions for physical work environments. The tool grapes the crews demand rather than what their work practices are in reality. While, the central concern with in co-design is to deal with the relation between studying the work practices of the workers from whom new technologies are being developed and directly engaging workers in design [9]. Thus, it is too dangerous to only utilise a piece of co-design and overlook another part. Furthermore, it is a challenge to conclude that there is a human-centred approach in the maritime studies [10–12] although a few researchers claim such concept elsewhere.

### 4.1 The focus of designing simulator-based maritime research

What is co-design about? According to Blomberg and Karasti[13]:

> The approach[Authors' interpretation] has been defined by its insistence that worker's knowledge is available to shape design directions by providing places and spaces for interaction between designers and practitioners that do not privilege one kind of knowledge over another.

The approach brings unique experiences and perspectives when people mutual learn from others' domain of knowledge. Everyone who participates in the design process has a voice that can be heard and be considered during the design process. This is a vital point for controlling the quality of a research and development project. With the increased concerns of safety maritime operation, designers are pushed to seek most appropriate approach to deal with such interests. However, we have to warn that it might be good to make visible participants' situated methods for creating the coherence of phenomena, such as applying the studied results from ergonomists regarding the traditional engineering design work, however, we lose the opportunity to describe phenomena using participants' categories and organising frameworks.

Due to non-existing systematic approach in the maritime domain, one could not find in-depth discussions regarding how technology can be and should be implemented in the maritime domain. Co-design can bring changes that is defined by the interests of workers, the requirements for their work, and the jointly negotiated path to change. Although researchers, developers, managers and others in the maritime domain might have different expertise and favourite in their own fields, they could find their ways to make the project more sustainable. As Bødker et al (Bødker, Kensing, & Simonsen, 2004, pp.140-141) remarked:

> Good IT design requires knowledge of work practices in order to determine which company traditions are fundamental and sustainable, and which are outdated. Put in a different way, only when a design team has fundamental knowledge of existing work practices can it arrive at what we call a 'sustainable design'.

In this case, all participants are the actors to shape the future in the maritime education. Maritime education may no longer only about engineering, electrician, management, and training, it becomes complex and with less clear boundary with other courses. That means everyone becomes co-designer and must opportunities to see first-hand, participant in, the life of the user participants. This is essential for the maritime education for the future skills. What competence should one to have in the digital era?

## 4.2 The change for the simulation-based maritime education

In order to better prepare for the future, we need to include studying phenomena in a systematic way of participants in their everyday settings, taking a holistic view, providing a descriptive understanding, and taking a member's perspective. Therefore, there is no necessary to distinguish who is providing what types of maritime education, but we can see them as a completely organisational system, including humans, technological artefacts, and institutional rules for organising humans and technologies together.

The starting point is always to find a way of providing socially enriched understanding of current work practices that is fruitful for designing simulation-based maritime education. It is firstly important to respect for the different knowledge that seafarers, engineers, technician, manager, and designers bring to the project. In this manner, we could commitment to a members' perspective that focuses on gaining an insider's view and using terms relevant and meaningful to the people who use simulators. This is the best way to create opportunities for designers and workers to learn about each other's domain through direct interaction for co-creating situations where seafarers can experience the design possibilities and encounter first-hand experiences. Secondly, it is also important to have a holistic view of how the outcome of the design that may affect the work practices of all participants. For example, changes in creating a scenario of maritime training that may request an impact on the engineering, design, teaching as well as management skills. Thirdly, describing current situation is important to prescribing a change. This is because without better knowing current situation is a vial resource to anchor change in the past and present, and offering all participants a limit scheme for the future imagination. Fourthly, since everyone is participating in designing scenarios-based maritime education, everyone is co-designer and must have opportunities to see first-hand, participate in, the life of the maritime education. The participatory designer, in this unique situation, can engage in a continuum of 'roles' with the ability to cycle between participation in the life of all simulator users and looking for new possibilities for changes.

## 4.3 Education providers as mediators

We need to stress that educational providers are mediators between the workplace and the design intervention for simulation-based maritime education. Simulators are not only products one developed for others to use. Also, simulators are not one who can only use for teaching purposes. We must acknowledge that simulation is only a tool that is used to support human's cooperation, collaboration, and maybe competition. However, without mutual learning process, we cannot confidently state that the non-transferable skills of different experts in their own fields can be grounded firmly via simulation-based maritime education. Thus, it will be a challenge for using simulator as a tool to promise educational goals, including training for the future.

In such consequence, education providers have to shift their positions from only providers' position to the positions of mediators. In tradition, educational providers only provide either educating people to design technology, or training people to use technology. This single way of education cannot promise simulator-based maritime education will help seafarers to be professionals; neither can help other participants have a clear and complete direction of maritime development. This is understandable that maritime domain was and is following the development of normal science [15], following the cognitive processor (procedure learning) [16] rather taking humans learning into account, which might base on intimated knowledge of several thousand concrete cases in peoples own area of expertise [5]. Co-design has contributed to change and offered an approach to help linking back the knowledge of work practices to the design of technology. In this process, co-design can help avoiding useless repetition of evaluation of training results, not to say its limitation of identifying one's true expertise. Instead, PD shifts our focus how to bring that expertise in the cycle to design. Winograd and Flores[17] add:

> We encounter the deep questions of design when we recognize that in designing tools, we are designing ways of being.

Design is, fundamentally for us, about designing futures for actual people. If people wish to encounter digitalisation, autonomous and other attractive activities in digitalisation era, we must agree that it is simulation-based maritime education is a system where co-design can facility different techniques to make innovation for the maritime domain, especially focusing on the future skills and competence in the digital future. Co-design is valuable in making visible 'multiple communities' in the maritime studies and do not leave 'distance area' for unmeasurable expertise in the design process. Instead, co-design allows creating a disciplinary division of labour, the differing expertise complementing one another. In this view, interdisciplinary is seen as a functional activity can be viewed as seeking its own ways of representing 'methodological' positions of different fields to work in a common place for making innovation.

## 5     Concluding Remarks

Although this is a point of departure for discussing how co-design can play a role to develop a methodology for foresight of future skills in the maritime domain, we find there is huge potential to restructure maritime education and research as a basis to support foreseeing competence of maritime personnel. In the article we argue that using a bottom-to-top design process to forecast human capability we can shape key features of future skills, as well as processes of linking past and current skills and knowledge to future needs. Through the process of the co-design approach, many participants from industry, research institutions, training companies, and authorities could cooperatively offer valuable insights into structuring the future. The most re-quired of us is to deploy this approach speciously into practice to improve simulation-based maritime education and training for the benefit of the future maritime labour force.